\documentclass[runningheads]{llncs}
\usepackage[T1]{fontenc}
\usepackage{graphicx}
\usepackage{amsmath}
\usepackage{amssymb}
\usepackage{mathtools}
\usepackage{hyperref}
\usepackage{cleveref}

\usepackage{booktabs}

\usepackage{algorithm}
\usepackage{algpseudocode}
\algnewcommand\algorithmicforeach{\textbf{for each}}
\algdef{S}[FOR]{ForEach}[1]{\algorithmicforeach\ #1\ \algorithmicdo}

\begin{document}
\title{Sphere of Influence Centrality via Shapley Values: Empirical Approximation and Network Coverage Analysis}
%
%
\author{Sie Hendrata Dharmawan\inst{1}\orcidID{0009-0008-5078-9907} \and \\
Kevin Limanta\inst{1}\orcidID{0000-0002-0098-7373} \and \\
Brendan Liu\inst{1}\orcidID{0009-0002-1683-4862} \and \\ 
Peter Chin\inst{1}\orcidID{0000-0002-1913-4223}}
%
\authorrunning{S. Dharmawan et al.}
%
\institute{Dartmouth College, Hanover, NH 03755 \\
\email{\{sie.hendrata.dharmawan.th, klimanta.th, brendan.liu.th, peter.chin\}@dartmouth.edu}\\}
%
\maketitle              
\begin{abstract}
Node centrality is a fundamental problem in network analysis, yet classical metrics fail to capture the collective, coalitional nature of influence. We present a systematic empirical evaluation of the Shapley-value-based framework for the sphere of influence problem --- selecting $m$ nodes to maximize network coverage under three reachability criteria: single-hop, $k$-hop, and multi-path connectivity --- using exact polynomial-time algorithms due to Michalak et al. Evaluation across three diverse real-world networks (Euroroad, Facebook TV Shows, and Cora) demonstrates that practical approximation ratios consistently approach 0.9, substantially exceeding the theoretical $(1-1/e)$ lower bound, and that the Shapley-based approach dramatically outperforms a degree-based baseline, particularly in hub-and-spoke topologies. In the most striking case, Shapley-based selection identifies just 26 nodes (under 1\% of the Cora network) sufficient to influence half the graph under 3-hop reachability, compared to substantially larger sets required by the naive baseline.

\end{abstract}

\section{Introduction}
Identifying the most strategically important nodes in a network is a central problem across many domains. Classical centrality measures such as degree, betweenness, and closeness quantify individual node importance efficiently, but they treat nodes in isolation. In practice, influence is a collective phenomenon: the combined effect of a set of nodes can differ substantially from the sum of their individual contributions, and a node that appears unremarkable by degree may be critical when acting alongside others.

This paper studies the \textbf{sphere of influence} problem, which formalizes this collective perspective. Given a graph $G$ and a budget $m$, the goal is to select a coalition $S$ of $m$ nodes whose sphere of influence (the set of nodes covered by $S$ under a chosen reachability criterion) is maximized. We consider three reachability criteria introduced in prior work \cite{Suri,Michalak}: coverage within 1-hop distance ($g_{\text{base}}$), coverage within $k$-hop distance ($g_{\text{hop}}$), and coverage by at least $k$ distinct coalition members within 1-hop distance ($g_{\text{conn}}$). All three variants are NP-hard \cite{Suri}, making approximation algorithms essential.

The approximation technique we build on leverages Shapley values from cooperative game theory \cite{P-295}. Each node's Shapley value captures its expected marginal contribution to coverage across all possible coalition orderings, satisfying the axioms of efficiency, symmetry, linearity, and the null player property \cite{WINTER20022025}. Selecting the $m$ nodes with highest Shapley values yields a $(1-1/e) \approx 0.63$ approximation guarantee \cite{Suri}. Crucially, Michalak et al. \cite{Michalak} showed that exact Shapley values for all three game variants can be computed in polynomial time, eliminating the Monte Carlo estimation required by the original approach.

While the theoretical foundations are well established, the \textbf{practical} performance of this framework across diverse real-world network topologies has not been systematically studied. The contributions of this paper are as follows:
\begin{enumerate}
    \item A comprehensive empirical evaluation of the Shapley-value-based framework across three structurally diverse real-world networks (Euroroad, Facebook TV Shows, and Cora), demonstrating that practical approximation ratios consistently approach 0.9, well above the theoretical $(1-1/e)$ floor.
    \item An empirical comparison against a degree-based baseline showing that Shapley-based selection substantially outperforms naive heuristics in heterogeneous network topologies, with the performance gap most pronounced in hub-and-spoke structures.
    \item A systematic analysis of the qualitative differences between the three reachability models ($g_{\text{base}}$, $g_{\text{hop}}$, $g_{\text{conn}}$), revealing a fundamental trade-off between reachability-based and consensus-based influence, and demonstrating remarkable efficiency: as few as 2 nodes suffice to cover 50\% of the Facebook network at $k=3$, and 26 nodes suffice in Cora.
\end{enumerate}

The paper is structured as follows. Section \ref{section:related} surveys related work. Section \ref{section:definition} defines notation and the three game formulations with their exact Shapley value algorithms. Section \ref{section:correctness} empirically verifies the approximation ratio on random graphs. Section \ref{section:datasets} describes the real-world datasets. Section \ref{section:experimental} presents experimental results. Section \ref{section:conclusion} presents our conclusion and future directions.

\section{Related Work}
\label{section:related}
\textbf{Node Centrality.} A broad survey of centrality measures in complex networks is provided by Saxena and Iyengar \cite{saxena2020centralitymeasurescomplexnetworks}, covering degree, betweenness, closeness, and eigenvector-based metrics including PageRank. These measures are efficiently computable but capture only individual structural properties. Our work departs from this tradition by adopting a cooperative game-theoretic formulation that models nodes' contributions when acting collectively — a distinction with direct consequences for which nodes are selected and how much coverage they achieve.

\textbf{Influence Maximization.} Kempe, Kleinberg, and Tardos \cite{Kempe2003MaximizingTS} formally introduced the influence maximization problem, proving NP-hardness under probabilistic cascade models and establishing the $(1-1/e)$ greedy approximation guarantee via submodularity. Subsequent work by Leskovec et al. \cite{leskovec} introduced the CELF algorithm, significantly reducing the Monte Carlo simulation cost of evaluating marginal gains. Our sphere of influence formulation shares the same approximation structure but adopts deterministic reachability criteria rather than probabilistic diffusion, which enables exact, closed-form Shapley value computation without simulation.

\textbf{Game-Theoretic Network Centrality.} Suri and Narahari \cite{Suri} pioneered the coalitional game formulation of the sphere of influence problem and established the $(1-1/e)$ approximation guarantee for Shapley-based node selection, relying on Monte Carlo estimation of Shapley values. Michalak et al. \cite{Michalak} significantly advanced this by deriving polynomial-time exact algorithms for the base, $k$-hop, and connectivity variants — the algorithms we implement and evaluate here. Narayanam and Narahari \cite{5499450} explored Shapley values for influence in network formation games, further demonstrating the breadth of this game-theoretic lens.

\textbf{Submodular Optimization.} The $(1-1/e)$ guarantee common to all three game variants follows from the classical result of Nemhauser, Wolsey, and Fisher \cite{Nemhauser} on greedy maximization of monotone submodular functions under cardinality constraints. Feige \cite{feige} established that this ratio cannot be improved beyond $(1-1/e)$ unless P = NP, contextualizing our empirical finding that practical ratios near 0.9 represent a meaningful gap above the worst-case bound.

\textbf{Domain-Specific Network Analysis.} The three datasets used in our evaluation (Euroroad \cite{nr}, Facebook TV Shows \cite{rozemberczki2019gemsec}, and Cora \cite{bojchevski2018deepgaussianembeddinggraphs}) represent qualitatively distinct network topologies: spatially constrained infrastructure networks, community-structured social networks, and hierarchically influenced citation networks, respectively. These structural differences make them collectively well-suited to stress-testing whether the advantages of Shapley-based selection generalize beyond homogeneous random graphs, which is a central empirical question of this paper.

\section{Definition And Notation} \label{section:definition}

In this section we reuse the notation from \cite{Michalak}. A \textit{graph} $G$ consists of \textit{nodes} and \textit{edges}, sets of which will be denoted $V(G)$ and $E(G)$ respectively. Every edge from set $E(G)$ connects two nodes in set $V(G)$. An edge $(u,v)$ connects vertices $u,v \in V(G)$. The number of edges incident to a vertex is called a \textit{vertex degree}. The \textit{neighboring vertices} of $v \in V$ are all vertices connected to $v$. A path is, informally, a sequence of connected edges. 

A \textit{coalition} $S = \{v_1, \dots, v_k\} \subset V(G)$ represents a subset of nodes within graph $G$. We define a \textit{characteristic function} $\nu(S)$ that maps any coalition to another coalition, with the corresponding payoff measured by $|\nu(S)|$ (the cardinality of the resulting coalition). Across all three game formulations presented below, we maintain the property $\nu(V(G)) = V(G)$, indicating that a coalition encompassing all graph vertices yields the entire vertex set as payoff. This formulation naturally enables the computation of Shapley values, where each node's importance is quantified as its expected marginal contribution when joining randomly formed coalitions. The mathematical specificity of each game variant is defined as follows

\subsection{ Game 1: Single-Hop Influence ($g_{base}$)}

This game was first introduced by \cite{Suri}. The payoff function is defined as the set of nodes that are at most 1 hop away, thereby including both the coalition $S$ itself as well as all immediate neighbors of nodes in $S$. \cite{Suri} established that computing the optimal solution for game $g_{base}$ is NP-hard, but crucially demonstrated that a Shapley-value approach provides a $(1-1/e)$ approximation guarantee of the optimal solution. Specifically, if $S$ comprises the top $m$ nodes with highest Shapley values, then:

$$|\nu(S)| \geq \left(1 - \frac{1}{e} \right) |\nu(S')|$$

where $S'$ represents the optimal solution, which would otherwise require brute-force computation. While \cite{Suri} employed a Monte-Carlo simulation approach to estimate Shapley values for this coalitional game, \cite{Michalak} subsequently derived Algorithm \ref{algo:gbase}, which computes exact Shapley values efficiently and provided formal proof of its correctness.

\begin{algorithm} 
\caption{Computing the Shapley Value for $g_{base}$}
\label{algo:gbase}
\begin{algorithmic}[1]
\ForEach {$v \in V(G) $}
    \State $SV[v] = \frac{1}{1 + deg_G(v)}$
    \ForEach{$u \in N_G(v)$}
        \State $SV[v] += \frac{1}{1 + deg_G(u)}$
    \EndFor
\EndFor
\State return SV
\end{algorithmic}
\end{algorithm}

\subsection{Game 2: $k$-Hop Influence ($g_{hop}$)}

This game represents a generalization of $g_{base}$ by extending the influence radius to include nodes that are at most $k$ hops away from any node in the coalition. Since $g_{base}$ is proven to be NP-hard, it follows by reduction that $g_{hop}$ is also NP-hard. Nevertheless, \cite{Michalak} developed Algorithm \ref{algo:ghop} for efficiently calculating the exact Shapley values for this extended influence model and provided formal verification of its correctness. The theoretical guarantee that this algorithm achieves a $(1-1/e)$ approximation ratio follows analogously from the proof framework established by \cite{Suri} for the single-hop case, as the underlying submodularity properties are preserved when extending from immediate neighbors to $k$-hop neighborhoods.

\begin{algorithm}
\caption{Computing the Shapley Value for $g_{hop}$}
\label{algo:ghop}
\begin{algorithmic}[1]
\ForEach {$v \in V(G)$}
    \State DistanceVector $D \gets \textsc{Dijkstra}(v, G)$
    \State $\text{extNeighbors}(v) \gets \{ u \in V(G) \mid D[u] \le k \}$
    \State $\text{extDegree}(v) \gets |\text{extNeighbors}(v)|$
\EndFor
\ForEach {$v \in V(G)$}
    \State $SV[v] \gets \frac{1}{1 + \text{extDegree}(v)}$
    \ForEach {$u \in \text{extNeighbors}(v)$}
        \State $SV[v] \gets SV[v] + \frac{1}{1 + \text{extDegree}(u)}$
    \EndFor
\EndFor
\State \textbf{return} $SV$
\end{algorithmic}
\end{algorithm}

\subsection{Game 3: Multiple-Path Influence ($g_{conn}$)}

This game introduces an alternative generalization of $g_{base}$ by implementing a threshold criterion rather than extending distance. The payoff function includes nodes that are reachable by at least $k$ different members of the coalition within 1-hop distance. Given the NP-hardness of $ g_{base} $, it follows by reduction that $g_{conn}$ is also computationally intractable. To address this challenge, \cite{Michalak} derived Algorithm \ref{algo:g3}, which efficiently calculates exact Shapley values for this threshold-based influence model and provided rigorous verification of its correctness. The theoretical guarantee that this selection method achieves a $(1-1/e)$ approximation ratio follows directly from the analytical framework established by \cite{Suri} for the base case, as the critical submodularity properties are preserved when transitioning from single-path influence to multiple-path influence requirements.

\begin{algorithm} \caption{Computing the Shapley Value for $g_{conn}$}
\label{algo:g3}
\begin{algorithmic}[1]
\ForEach{$v \in V(G)$}
    \State SV[$v$] $= \frac{1}{1+\text{degG}(v)}$
    \ForEach{$u \in N_G(v)$}
        \State SV[$v$] $+= \max{(0, \frac{(\text{degG}(u)-k+1)}{\text{degG}(u)(1+\text{degG}(u))})}$
    \EndFor
\EndFor
\State return SV
\end{algorithmic}
\end{algorithm}

\subsection{Scalability Analysis}

The three algorithms differ substantially in their scalability to large networks. Algorithms \ref{algo:gbase} and \ref{algo:g3} both consist of a single pass over all nodes and their immediate neighbors, yielding $O(|V| + |E|)$ time complexity --- linear in the size of the graph and thus readily applicable to networks with millions of nodes. Algorithm \ref{algo:ghop} is more expensive: it runs Dijkstra's algorithm from every node to compute $k$-hop neighborhoods, giving $O(|V| \cdot |E| \log |V|)$ complexity, which becomes the primary computational bottleneck for large dense graphs. In practice, this cost can be mitigated by pruning the search at depth $k$ rather than computing full shortest-path trees, reducing the effective neighborhood explored per node. The brute-force optimal used in Section \ref{section:correctness} is exponential in $m$ and is only feasible for the small random graphs ($n = 50$--$100$) used in that evaluation; it is not intended for deployment. For large-scale applications, the Shapley-based algorithms --- particularly $g_{\text{base}}$ and $g_{\text{conn}}$ --- offer an attractive combination of polynomial-time computation and strong empirical approximation guarantees.

\section{Empirical Verification of Approximation Ratio} \label{section:correctness}

While \cite{Suri} established the theoretical guarantee that the Shapley-value-based algorithm achieves a $(1-1/e) \approx 0.63$ approximation of the optimal solution for these NP-hard problems, this ratio appears relatively modest for practical applications. To assess real-world performance, we conducted extensive experimentation on randomly generated graph instances. Our empirical results demonstrate that the typical approximation ratio substantially exceeds the theoretical lower bound, consistently achieving values closer to 0.9. This improved performance significantly enhances the practical viability of our approach across diverse application domains, as stakeholders in resource allocation, influence maximization, and network security contexts generally require higher approximation guarantees to justify implementation.

\begin{figure}[h!]
    \centering
    \begin{minipage}{0.25\textwidth} 
        \centering
        \includegraphics[width=\linewidth]{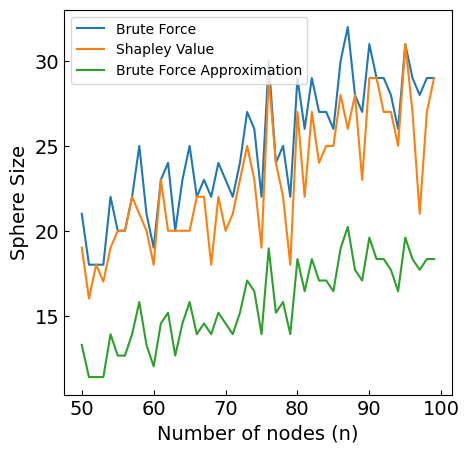}
    \end{minipage}
    \begin{minipage}{0.25\textwidth}
        \centering
        \includegraphics[width=\linewidth]{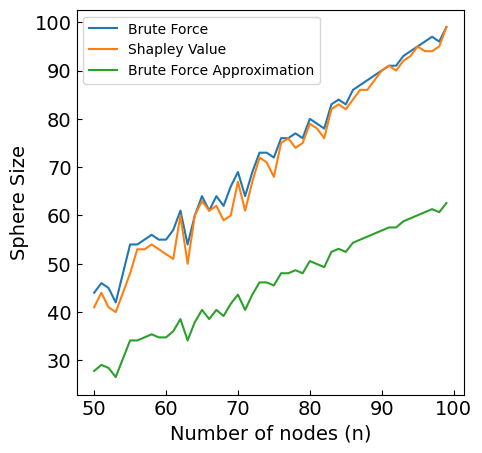}
    \end{minipage}
    \begin{minipage}{0.25\textwidth}
        \centering
        \includegraphics[width=\linewidth]{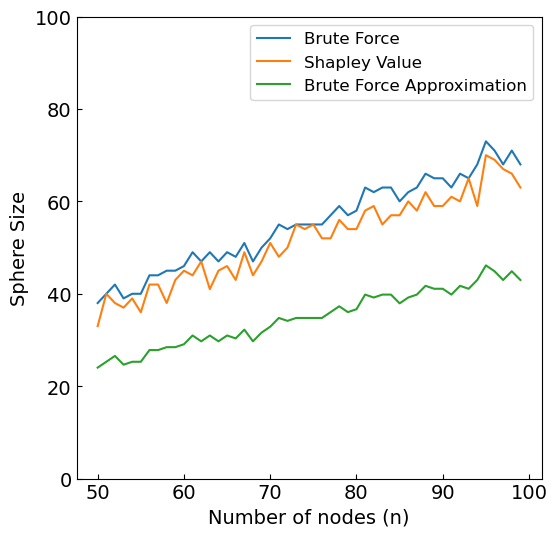}
    \end{minipage}
    \hfill
    \begin{minipage}{0.25\textwidth}
        \centering
        \includegraphics[width=0.9\linewidth]{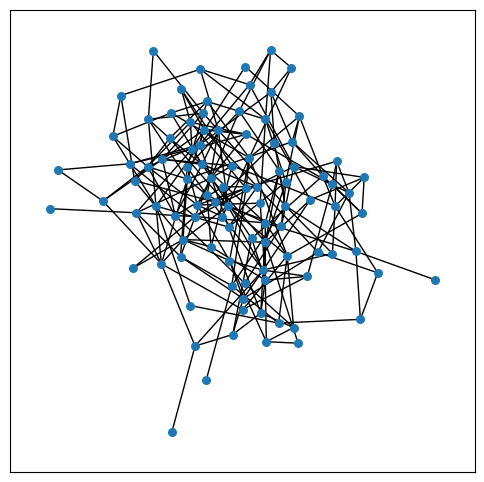}
    \end{minipage}
    \begin{minipage}{0.25\textwidth}
        \centering
        \includegraphics[width=0.9\linewidth]{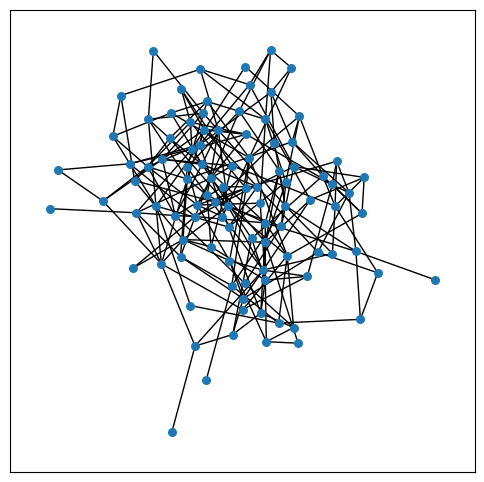}
    \end{minipage}    
    \begin{minipage}{0.25\textwidth}
        \centering
        \includegraphics[width=0.9\linewidth]{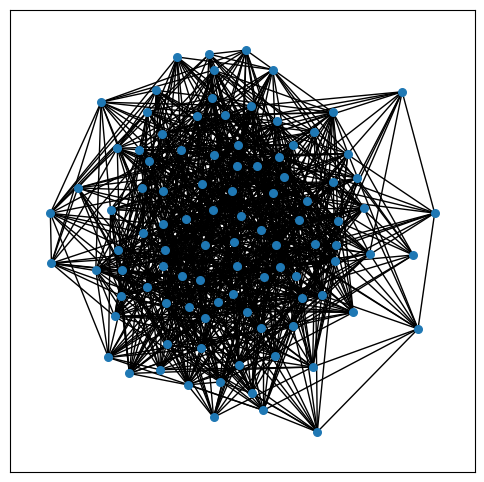}
    \end{minipage}
    \caption{Empirical verification of the $(1-1/e)$ approximation guarantee on Erdős–Rényi random graphs ($n = 50$–$100$, $p = 0.05$), shown for $g_{\text{base}}$ (left), $g_{\text{hop}}$ (center), and $g_{\text{conn}}$ (right) with $m = 3$ and $k = 2$. Above: blue lines show brute-force optimal sphere of influence size, orange lines show Shapley-based approximation results, and green lines mark the $(1-1/e)$ theoretical lower bound. The Shapley-based trajectories consistently exceed the lower bound and track closely to the optimum, with average approximation ratios of 0.93, 0.92, and 0.93 for $g_{\text{base}}$, $g_{\text{hop}}$, and $g_{\text{conn}}$ respectively. Below: representative graph instances illustrating the sparse, irregular structure typical of the test instances.}
    \label{fig:correctness}
\end{figure}

Our experimental validation utilized randomly generated graphs with vertex counts ranging from $n=50$ to $n=100$, where each potential edge was instantiated with probability $p=0.05$. Across all three game formulations, we fixed the selection parameter at $m=3$, seeking to identify a trio of nodes maximizing sphere of influence. For game $g_{hop}$, we employed a hop distance parameter $k=2$, and similarly for game $g_{conn}$, we implemented an influence threshold of $k=2$. The comparative results are illustrated in Figure \ref{fig:correctness}.

The horizontal axis represents graph order (node count), while the vertical axis quantifies sphere of influence magnitude. Blue lines indicate optimal solutions derived via exhaustive brute-force computation, orange lines represent our Shapley-based approximation results, and green lines demarcate the $(1-1/e)$ theoretical lower bound of optimal solutions. The empirical data demonstrate that orange lines consistently exceed green lines across all parameter configurations, empirically validating the theoretical $(1-1/e)$ approximation guarantee. More significantly, the orange lines exhibit substantially closer proximity to optimal blue lines than to the theoretical lower bound. Quantitatively, games $g_{base}$ and $g_{conn}$ achieve a minimum approximation ratio of 0.81 with an average ratio of 0.93, while game $g_{hop}$ exhibits a minimum ratio of 0.78 with an average ratio of 0.92. These minimum ratios are attributable to isolated pathological graph configurations, as evidenced by the average ratios approaching 0.9 across all formulations.

\begin{figure}[h!] 
    \centering
        \includegraphics[width=0.4\textwidth]{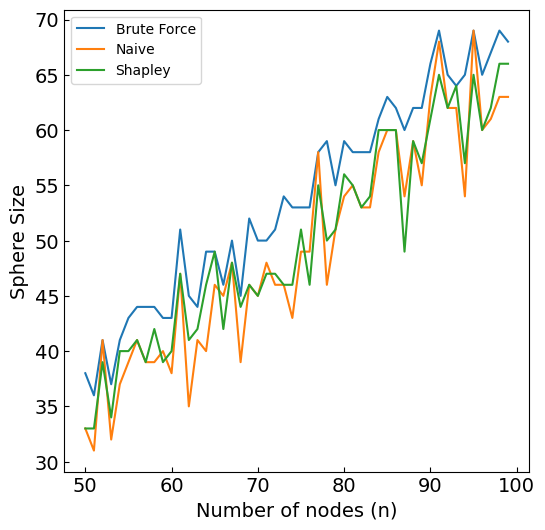}
    \caption{Sphere of influence size on random graphs ($n = 50$–$100$, $p = 0.05$) comparing Shapley-based selection (green), naive degree-based selection (orange), and brute-force optimum (blue) under $g_{\text{base}}$. The two approximation methods perform comparably under homogeneous edge density, with no consistent winner — a consequence of the structural uniformity of random graphs that obscures the coalitional advantages captured by Shapley values. This motivates the evaluation on structured real-world networks in Section 6, where heterogeneous topology makes the distinction between the two methods substantial.}
    \label{fig:sv_naive}
\end{figure}

In addition to the brute force method and our Shapley value-based approach, we implemented a naive baseline for $g_{base}$: selecting the $m$ nodes with highest degrees. This baseline is motivated by the observation that high-centrality nodes often exhibit high degrees in practice. Figure \ref{fig:sv_naive} compares the sphere of influence results from our Shapley value method against both the naive approach and brute force (included as ground truth). 

On random graphs with uniform edge density, we observe that the naive method and Shapley value approach perform comparably, with no clear winner. This is expected given the homogeneous structure of such graphs. However, as we demonstrate in subsequent sections using real-world datasets, the Shapley value-based method significantly outperforms the naive baseline, suggesting that our game-theoretic framework captures important structural properties beyond simple degree information.

\section{Datasets}
\label{section:datasets}

For experiments on real datasets, we selected three networks representing diverse application domains:

\begin{enumerate}
    \item \textbf{Euroroad} \cite{nr}: A transportation network of European roads and cities, where nodes represent cities and edges represent direct road connections. This network exhibits spatial constraints and hub-like structures around major metropolitan areas.
    
    \item \textbf{Facebook TV Shows} \cite{rozemberczki2019gemsec}: A social network where nodes represent Facebook pages about TV shows and edges indicate mutual likes between pages. This captures community structures and preference patterns in social media.
    
    \item \textbf{Cora} \cite{bojchevski2018deepgaussianembeddinggraphs}: A citation network of machine learning papers, where edges represent citations. This academic network demonstrates hierarchical influence patterns and research community structures.
\end{enumerate}

\section{Experimental Results}
\label{section:experimental}

Across all three game formulations and datasets, the Shapley-based approach consistently outperforms the naive degree-based baseline. Table \ref{tab:coalition_sizes} summarizes the coalition sizes required to reach 50\% network coverage under each method, providing a compact view of the performance gap before the subsections discuss each game in detail.

\begin{table}[h]
\centering
\caption{Coalition size required to reach 50\% network coverage under each method and game formulation.
Shapley-based selection is used for all games; Naive refers to degree-based selection under $g_{\text{base}}$.}
\label{tab:coalition_sizes}
\begin{tabular}{lcccc}
\toprule
\textbf{Dataset} 
    & \textbf{$g_{\text{base}}$ } 
    & \textbf{$g_{\text{base}}$ (Naive)} 
    & \textbf{$g_{\text{hop}}$ ($k=3$) } 
    & \textbf{$g_{\text{conn}}$ ($k=2$) } \\
\midrule
Euroroad     & 134 & 233 & 131 & 544  \\
Facebook     & 134 & 286 &   2 & 902  \\
Cora         &  99 & 496 &  26 & 995  \\
\bottomrule
\end{tabular}
\end{table}

\subsection{Base game ($g_{\text{base}}$)}
\begin{figure}[ht!] 
    \centering
    \begin{minipage}{0.32\textwidth} 
        \centering
        \includegraphics[width=\linewidth]{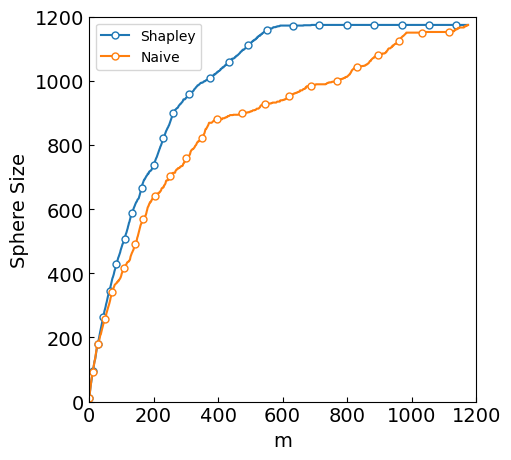}
    \end{minipage}
    \begin{minipage}{0.32\textwidth} 
        \centering
        \includegraphics[width=\linewidth]{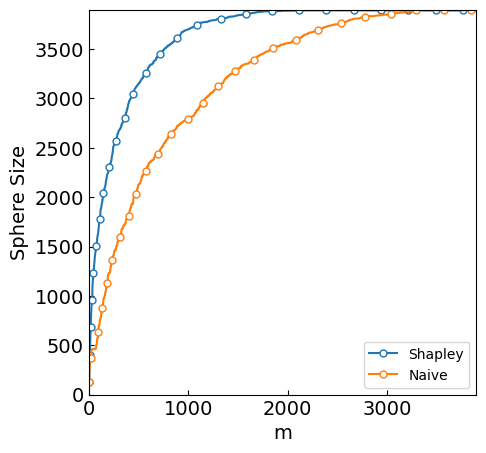}
    \end{minipage}
    \begin{minipage}{0.32\textwidth} 
        \centering
        \includegraphics[width=\linewidth]{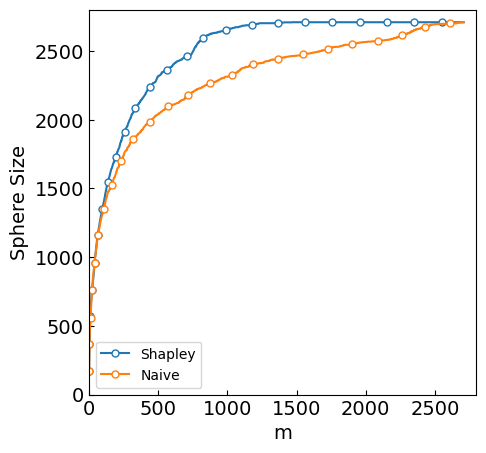}
    \end{minipage}
    \caption{Sphere of influence size as a function of coalition size $m$ under the base game ($g_{base}$) for Euroroad (left), Facebook TV Shows (center), and Cora (right). The Shapley-based method (blue) consistently dominates the naive degree-based baseline (orange) across all three networks, with the performance gap widening in networks with pronounced hub-and-spoke structure. To reach 50\% coverage, Shapley-based selection requires coalitions of $m=134,134,99$ nodes in Euroroad, Facebook, and Cora respectively, compared to substantially larger sets under the naive approach (see Table \ref{tab:coalition_sizes}).}
    \label{fig:g1}
\end{figure}

Figure \labelcref{fig:g1} shows sphere of influence size as a function of coalition size $m$ for the base game across all three datasets, with graph visualizations of the minimum coalition achieving 50\% coverage. The Shapley-based approach uniformly achieves larger spheres with fewer nodes than the naive degree-based baseline, but the magnitude of the advantage differs meaningfully across network types and illuminates what Shapley values capture that degree alone cannot.

In the Euroroad network, the gap is moderate but consistent. The network's spatial, hub-and-spoke structure means that high-degree nodes tend to be genuinely central, so the degree baseline is a reasonable heuristic. But Shapley values still identify a more efficient coalition by accounting for the collective coverage overlap between hub cities.

The performance gap widens substantially in the Facebook TV Shows and Cora networks. In Facebook, certain pages serve as cross-community connectors whose structural role is not fully reflected in their degree; in Cora, seminal papers act as citation hubs bridging disparate research areas. In both cases, Shapley values identify nodes whose influence cascades far beyond their immediate neighborhoods. The Cora result is particularly striking: just 99 nodes ($\sim$ 3.3\% of the network) suffice to reach 50\% coverage, compared to significantly more under the naive approach. This gap is a direct consequence of the Shapley framework's ability to account for the cascading, non-additive nature of influence in citation networks.

\subsection{$k$-hop game ($g_{\text{hop}}$)}

\begin{figure}[ht!] 
    \centering
    \begin{minipage}{0.32\textwidth} 
        \centering
        \includegraphics[width=\linewidth]{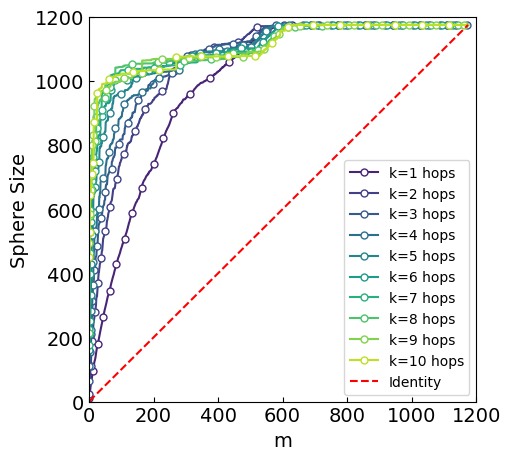}
    \end{minipage}
    \begin{minipage}{0.32\textwidth} 
        \centering
        \includegraphics[width=\linewidth]{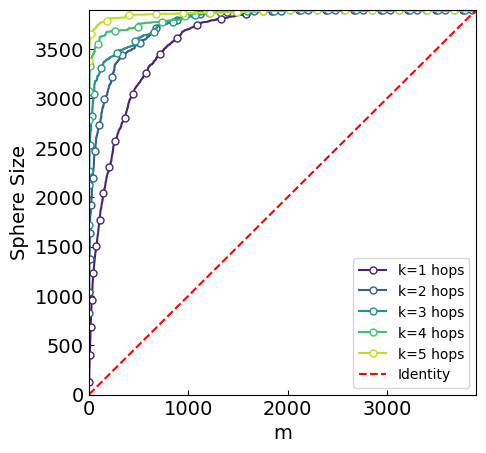}
    \end{minipage}
    \begin{minipage}{0.32\textwidth} 
        \centering
        \includegraphics[width=\linewidth]{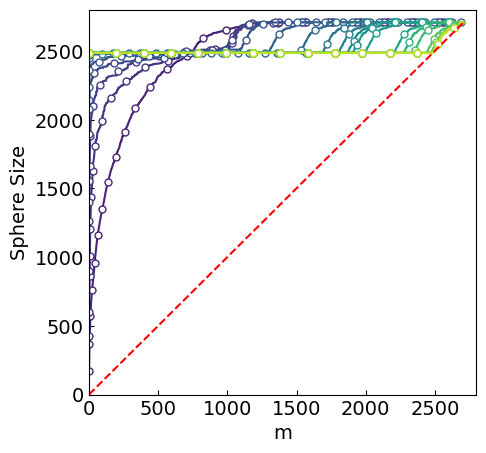}
    \end{minipage}
    \caption{Sphere of influence size as a function of coalition size $m$ under the $k$-hop game ($g_{hop}$) for Euroroad (left), Facebook TV Shows (center), and Cora (right), shown for $k=1,\dots,10$. The dotted red line indicates the identity $f(m)=m$. Increasing $k$ expands coverage monotonically in all networks, but the rate differs markedly by topology: Facebook's small-world structure enables 50\% coverage with just $m=2$ nodes at $k=3$, while Euroroad's spatial constraints slow coverage growth and produce a brief anomaly near $m=600$ where Shapley rankings optimized for smaller $k$ become suboptimal as the hop radius grows.}
    \label{fig:g2}
\end{figure}

Figure \labelcref{fig:g2} shows sphere of influence size for varying values of $k$ across all three datasets, with the identity line $f(m) = m$ as a reference. As expected, increasing $k$ monotonically expands coverage since every node's reachable set grows. The more revealing finding is how steeply coverage grows with $k$ in densely connected networks, and how small the required coalition becomes as a result.

The Facebook network exhibits the most dramatic effect: with only $m = 2$ nodes and $k = 3$, the method achieves 50\% coverage of the entire network. This reflects Facebook's small-world connectivity, where a handful of central pages can reach most of the graph within three hops. The Cora network requires slightly more (26 nodes at $k = 3$) but this still represents under 1\% of the network and is a substantial reduction from the 99 needed in the single-hop case.

The Euroroad network presents a different character. Coverage grows more slowly with $k$, consistent with its spatial embedding and low average degree. A notable anomaly appears near $m = 600$: the sphere of influence briefly decreases as $k$ increases from certain values. This occurs because the Shapley-based ranking is computed for a fixed $k$, so nodes selected as optimal seeds for smaller $k$ may reside in sparse peripheral regions that are less advantageous at larger hop distances, causing the aggregate sphere to shrink before recovering as more of the network becomes reachable.

\subsection{Multiple-path influence game ($g_{\text{conn}}$)}

\begin{figure}[ht!] 
    \centering
    \begin{minipage}{0.32\textwidth} 
        \centering
        \includegraphics[width=\linewidth]{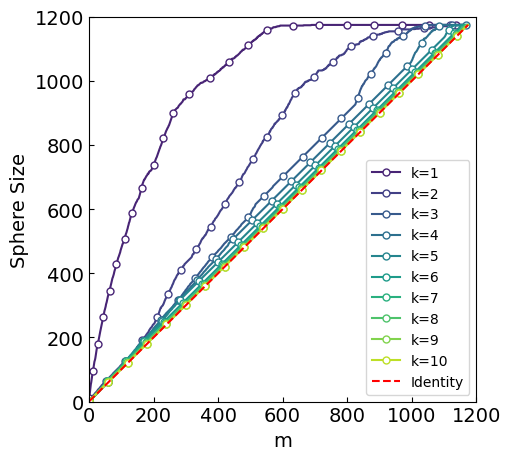}
    \end{minipage}
    \begin{minipage}{0.32\textwidth} 
        \centering
        \includegraphics[width=\linewidth]{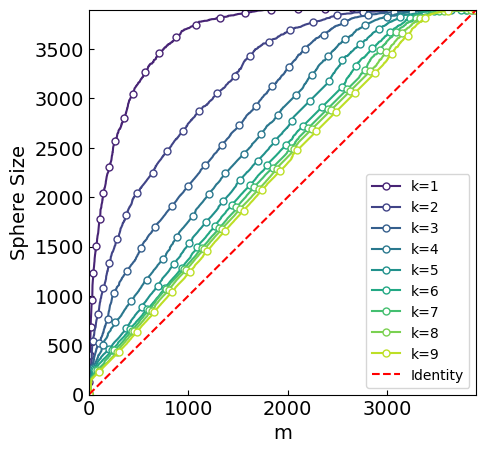}
    \end{minipage}
    \begin{minipage}{0.32\textwidth} 
        \centering
        \includegraphics[width=\linewidth]{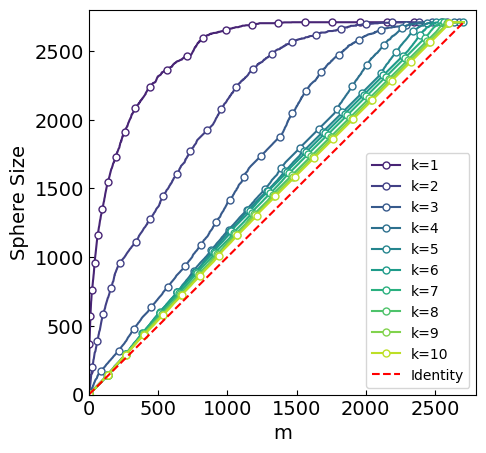}
    \end{minipage}
    \caption{Sphere of influence size as a function of coalition size $m$ under the connectivity game ($g_{conn}$) for Euroroad (left), Facebook TV Shows (center), and Cora (right), shown for $k=1,\dots,10$. The dotted red line indicates the identity $f(m)=m$. In contrast to $g_{hop}$, increasing $k$ shrinks the sphere of influence across all three networks, as nodes must satisfy the progressively stringent requirement of adjacency to at least $k$ coalition members. At high $k$ the sphere converges toward the identity line, indicating that most real-world networks lack the dense local connectivity required to sustain multi-path influence.}
    \label{fig:g3}
\end{figure}

The connectivity game $g_{\text{conn}}$ (Figure 5) reveals the most structurally distinct behavior of the three games. Unlike $g_{\text{hop}}$, where increasing $k$ expands coverage, here increasing $k$ shrinks it — because nodes must now be adjacent to at least $k$ coalition members simultaneously, a progressively harder condition to satisfy. This is directly visible in Figure 5: as $k$ increases, the sphere-size trajectories compress downward toward the identity line at different rates depending on network density, with Euroroad and Cora converging fastest due to their low to moderate average degrees, and Facebook slowing the collapse by virtue of its higher connectivity. Even so, Facebook cannot escape the constraint. The coalition size required for 50\% coverage under $g_{\text{conn}}$ at $k = 2$ is $m = 902$, an order of magnitude larger than the $m = 2$ sufficient under $g_{\text{hop}}$ at $k = 3$ (Table 1), underscoring the qualitatively different influence model each game encodes.

This trade-off between $g_{\text{hop}}$ and $g_{\text{conn}}$ is not merely a numerical curiosity. It reflects genuinely different application settings. The $k$-hop game models broad reachability under any propagation path, appropriate for information diffusion or logistics. The connectivity game models consensus or trust-based adoption, where a node is only influenced after receiving signals from multiple independent sources. The fact that most real-world networks struggle to sustain large spheres under $g_{\text{conn}}$ suggests that multi-path influence is an inherently constrained regime, making Shapley-based node selection all the more critical for identifying the rare nodes capable of sustaining it.

\section{Conclusion}
\label{section:conclusion}

This paper presented a systematic empirical study of Shapley-value-based node centrality under three sphere of influence formulations: single-hop, $k$-hop, and multi-path connectivity, across three structurally diverse real-world networks. Our results demonstrate that the practical approximation ratios achieved consistently approach 0.9, well above the theoretical $(1-1/e)$ floor, and that Shapley-based selection substantially outperforms degree-based heuristics wherever network structure departs from homogeneity. The performance advantage is most pronounced in hub-and-spoke topologies: in the Cora citation network, fewer than 1\% of the graph suffice to influence half the network at $k = 3$, a result that degree-based selection cannot approach. More broadly, the contrast between $g_{\text{hop}}$ and $g_{\text{conn}}$ reveals a fundamental trade-off between reachability-based and consensus-based influence models, with most real-world networks proving too sparsely connected to sustain large spheres under the latter.

Several natural extensions follow from this work. Incorporating heterogeneous edge weights and node values would allow the framework to model varying road capacities, relationship strengths, or paper impact rather than treating all nodes and edges as uniform. Developing incremental algorithms that update Shapley values as networks evolve would extend applicability to dynamic settings requiring real-time influence tracking. Finally, comparative studies against alternative solution concepts (the Banzhaf value, nucleolus, or core) could clarify when Shapley values are the most appropriate tool and when others offer advantages.

All code, including implementations of the three Shapley value algorithms, the degree-based baseline, and the experimental evaluation scripts for all datasets, is available on \url{https://github.com/hendrata-th/shapley-node-centrality}

\clearpage
%
%
%
%
\bibliographystyle{splncs04}
\bibliography{ref}
\newpage

\end{document}